# Unexpected Long-Term Variability in Jupiter's Tropospheric Temperatures

Glenn S. Orton[1*], Arrate Antuñano[2,3], Leigh N. Fletcher[3], James A. Sinclair[1], Thomas W. Momary[1], Takuya Fujiyoshi[4], Padma Yanamandra-Fisher[5], Padraig T. Donnelly[6], Jennifer J. Greco[7,8], Anna V. Payne[9], Kimberly A. Boydstun[10], Laura E. Wakefield[11.]

[1]Jet Propulsion Laboratory, California Institute of Technology, Pasadena, CA, USA
[2]Depertment of Applied Physics, University of the Basque Country, Bilbao, Spain
[3]School of Physics and Astronomy, University of Leicester, Leicester, UK
[4]Subaru Telescope, National Astronomical Observatory of Japan, National Institutes of Natural Sciences, Hilo, HI, USA
[5]Space Science Institute, Boulder, CO, USA
[6]Laboratoire de Meteorologie Dynamique, Institut Pierre-Simon Laplace, Sorbonne Université, Paris, France
[7]Science Department, Saint Ursula Academy, Toledo, OH, USA
[8]University of Hawaii, Honolulu, HI, USA
[9]Engineering - Betterlife, Mountain View, CA, USA
[10]Northrup-Grummon Corporation, San Diego, CA, USA

Abstract

An essential component of planetary climatology is knowledge of the tropospheric temperature field and its variability. Previous studies of Jupiter hinted at periodic behavior that was non-seasonal, as well as dynamical relationships between tropospheric and stratospheric temperatures. However, these observations were made over time frames shorter than Jupiter's orbit or they used sparse sampling. We derived upper-tropospheric (330-mbar) temperatures over 40 years, extending those studies to cover several orbits of Jupiter, revealing unexpected results. Periodicities of 4, 7, 8-9 and 10-14 years were discovered that involved different latitude bands and seem disconnected from seasonal changes in solar heating. Anticorrelations of variability in opposite hemispheres were particularly striking at 16°, 22° and 30° from the equator. Equatorial temperature variations are also anticorrelated with those 60-70 km above. Such behavior suggests a top-down control of equatorial tropospheric temperatures from stratospheric dynamics. Realistic future global climate models must address the origins of these variations in preparation for their extension to a wider array of gas-giant exoplanets.

Editor's Summary

Infrared observations of Jupiter obtained in a 40-yr timespace between 2918 and 2019 show long-term variations of tropospheric temperature with different periodicities, particularly at tropical latitudes, which often bear some connetion with sstratospheric temperature fluctuations.

## Main

Temperature and its variability constitute an essential element of the climatology of a planetary atmosphere, inextricably linked to planetary dynamics and chemistry. Initial studies of long-term variability of outer-planet temperatures focused on the stratosphere of Saturn[1,2,3,4]. Studies focused on Jupiter detected a stratospheric temperature oscillation[5,6] which – like Saturn - results from the vertical propagation of waves[7]. The earliest study of Jupiter's tropospheric temperatures examined 1- and 2-dimensional scans of its disk over 1980-1990[8], later extended in time to include 2-dimensional mid-infrared imaging through 2001[9]. Although snapshots of Jupiter's tropospheric temperatures have been investigated during epochs of visible change in Jupiter's banded features[10,11,12] there has been no systematic assessment of long-term tropospheric temperature variations.

## Observations

Here, we extend the record of Jupiter's infrared variability to cover 1978-2019, allowing us to diagnose Jupiter's upper-tropospheric temperature field for three and a half of its 11.9-year solar orbits, separating seasonal and non-seasonal variability unambiguously. The images were obtained using mid-infrared instruments mounted on NASA's Infrared Telescope Facility, the Subaru Telescope, and the Very Large Telescope, as described in the Methods section, the Materials section of the Supplemementary Information, and detailed individually in Supplementary Table 1. These observations were made using mid-infrared instruments with discrete filters at 17-25 µm, where the emission is sensitive to temperatures in the upper troposphere (100-400 mbar). In order to cover variability over such a long time frame, we chose a subset of these filters providing the longest continuous record, with central wavelengths of 18.72 and 20.50 µm. Some of the earliest observations that were made with only a broad filter near 18 µm between 1978 and 1983 (ref. 8) were analyzed independently, as discussed in the Supplementary Information. Figure 1 shows the 1983-2019 longitudinal-mean brightness temperatures of both filters. Figure 1c shows the retrieved 330-mbar temperatures and Figure 1d the residual from a longitudinal mean. Figure 2 illustrates temperatures at specific latitudes.

## Periodicities

Among the striking properties of the brightness-temperature time series (Fig. 1e, f) are apparent roughly 10-14 year periodicities of temperatures (the yellow, orange and red boxes in Fig. 1h), all close to Jupiter's 11.9-year orbital period. To ensure that this quasi-annual behavior is not an artifact of the seasonal variations of the emission angle at a fixed latitude, we estimated the size of the brightness temperature change due solely to the change of emission angle: even at 30° from the equator, the change (0.3 K) is much smaller than the observed variations. Furthermore, the apparent warmest and coldest temperatures are not coincident with the solstices, $L_s = 90°$ and 270° (see Figs. 2, Supplementary Figures 1 and 3), which would result from a purely geometric effect. In addition, the hemispheric temperature contrasts are consistent with those derived from independent studies of Voyager IRIS[13] and Cassini CIRS[14] observations (Fig. 2). The temperature variations are unlikely to result directly from radiative heating, as recent models[15] predict peak-to-peak tropospheric temperature seasonal variability of only 0.4 K or less, given Jupiter's small axial tilt. The models are also offset in time from the measurements (Supplementary Fig. 1). Time-dependent oscillations do exist in other planetary atmospheres that are related to seasonal cadences, such as Saturn's semi-annual equatorial oscillation[4], or perhaps more loosely, the Earth's 28-29 month Quasi-Biennial Oscillation (QBO)[16]. Both phenomena are tied to narrow low-latitude regions, which might imply similar mechanisms for Jupiter.

## Hemispherical asymmetries

However, such mechanisms do not explain one other striking characteristic: the variability of these temperatures is anticorrelated between the northern and southern hemispheres. This can be seen at the discrete latitudes shown in Figures 2b, c. Examining the Pearson correlation coefficient associated with conjugate latitudes yielded negative correlations over a broad range of latitudes, the largest at 16° (Fig. 3a), as well as 22° and 30° from the equator (Fig. 3b). This suggests teleconnected patterns of variability between the two hemispheres, such as the Earth's El Niño Southern Oscillation (ENSO)[17] and the North Atlantic Oscillation (NAO)[18] that are not well understood, and may well be related to one another. If Jupiter's quasi-annual tropospheric oscillations are driven from great depth, we would have expected that any anticorrelated patterns arising from connections via cylinders parallel to the rotation axis would be most effective equatorward of ±16° where the cylinders do not intersect with the inhibiting dynamics of a region of metallic hydrogen[19,20], but we observe exceptions to this at 22° and 30°. The anticorrelated variations could possibly be the result of seasonal variations of hazes that contribute substantially to upper-tropsopheric radiative balance. Stratospheric oscillations could also be modulating dynamic heating of the upper troposphere, possibly by controlling upward wave fluxes.

## Relation to stratospheric temperatures

A prominent period of 8.3±1.0 years, confined to ±10° of the equator, and a fainter period of 4.5±0.5 years are both detectable at 8°S-22°N (the blue and white boxes, respectively, in Fig. 1g). These tropospheric periodicities are similar to and may be related to the roughly 4-year equatorial stratospheric oscillation[21]. Temperature oscillations at 330 mbar (Fig. 2a) appear to be anticorrelated with equatorial zonal-mean stratospheric temperature oscillations in 1980-2011[21], which are consistent with the presence of the descending pattern of temperature anomalies detected in a study of the evolution of stratospheric temperatures at high vertical resolution, as shown in Fig. 3 of ref. 7. This implies a "top-down" control of tropospheric temperatures by the dynamics of the stratosphere, similar to "sudden warming" events in the Earth's atmosphere[22]. If both the 4.5- and 8.3-year periodicities are related to the equatorial stratospheric oscillation, the cause of the major difference in their latitudinal extent remains unresolved. Jupiter's zonal-mean winds have 13.8-year and 7.6-year variabilities[23], both confined to within 5° of the equator; the latter is near our low-latitude 8.3-year periodicity. A period of 7.4±0.5 years is also detectable over a wide range of latitudes (dark orange box in Fig 1h). Intriguingly, although apparently unrelated to variability in the major axisymmetric bands, a 7-year period is also detected in 5-μm studies near the equator[24], suggesting a possible correlation between temperature variability and the condensation of clouds in the 1-4 bar range.

## Comparison with other atmospheric properties

We found several correlations between temperatures and the visual appearance of Jupiter's prominent bands, looking particularly for correlations with dramatic quasi-periodic changes involving all longitudes of Jupiter's visibly dark belts and bright zones. Blue horizontal bars in Figure 2b denote epochs when the typically dark roughly 6-15°N North Equatorial Belt (NEB) expanded northward to cover 18.5°N[10,14], as illustrated by Supplementary Figure 3b. The periods of expansion in 2002, 2010 and 2018 appear to be coincident with prominent maxima of the NEB upper-tropospheric temperatures at 16°-30° N, consistent with the removal of aerosols by heating and sublimation. Equatorial Zone (EZ) disturbances[25,26] in 1992, 1999-2000 and 2006-2007, denoted by the black horizontal bars in Fig. 2a and illustrated by Supplementary Figure 3b, appear to be contemporaneous with decreases in 330-mbar temperatures, in general agreement with a detailed study of atmospheric properties for the 2006-2007 event[26]. This cadence is also consistent with the 6-7 year period of these events[25], even though not all of the equatorial temperature changes are tied to full-scale cloud-disturbance episodes. The red horizontal bars shown in Fig. 2b indicate the duration of South Equatorial Belt (SEB, 6°S-17°S) brightening and re-darkening ('fade' and 'revival') episodes, one of which is illustrated in Supplementary Figure 3c. All except a very brief sequence in the first half of 2007 coincide with the coldest periods at 16°S, which would be consistent with more aerosol condensation and visible whitening. On the other hand, periods when the prominent dark North Temperate Belt (NTB, roughly 22°-24°N, also identified in Supplementary Fig. 3) underwent lengthy brightening and re-darkening episodes related to spectacular plume activity (not shown in Fig. 2) appear to have no correlation with temperatures or their variability, at least at the resolution of our observations. We found no robust evidence for anticorrelations at conjugate latitudes north and south of the equator in the visible record that mirror those for 330-mbar temperatures shown in Fig. 3b. We note that previous investigations suggest that changes in temperatures and aerosols within a band do not occur at all longitudes simultaneously, but rather start in localised domains and spread with longitude over a finite time period, of order weeks[10,12]. For the purposes of this study, which probes timescales of years, we assume that the observations on individual dates are representative of the mean within a band at a certain time, but we note that this will lead to uncertainties in brightness temperature of ±1-2 K, as shown in Fig. 4 for the individual measurements. Figure 5 compares the 330-mbar zonal-mean temperatures with brightness temperatures at 7.9 µm, corresponding to stratospheric emission, which serve as a proxy for temperatures in Jupiter's stratosphere near 1-10 mbars of pressure, some 60-70 km higher in the atmosphere. We note that the anticorrelation with the 330-mbar temperatures at the equator is consistent with a top-down control of upper tropospheric temperature variations, i.e., the downward propagation of temperature anomalies associated with Jupiter's equatorial stratospheric oscillation[21].

## Conclusions

Our study of long-term zonal-mean tropospheric temperature variability in Jupiter has yielded evidence for both non-seasonal and quasi-seasonal periodicities at temperate to tropical latitudes along with associated puzzles. Although intriguingly close to Jupiter's 11.9-year orbit, the distinct 10-14-year periodicities are unlikely to be the direct result of radiative forcing in view of Jupiter's weak seasons, particularly given the pronounced hemispherical asymmetry of temperatures peaking at 16°, 22° and 30° from the equator. Such an asymmetry is mirrored in cloud opacity detected by 5-µm imaging, most prominently between the NEB and SEB[26]. The presence of 4- and 8-9-year periodicities suggests a relationship with stratospheric temperature variability. More detailed correlation between their periodicity and phase is needed to validate that connection, particularly to test the suggestion of "top-down" mechanisms, such as descending waves. The 7-year periodicity over a broad latitude range may also be related to the quasi-periodic equatorial disruption with the same cadence[26]. Although we found no straightforward correlations between periods and the latitudes of Jupiter's belts and zones, correlations with known global-scale changes of cloud morphology suggest at least some thermal modulation of aerosol condensation and sublimation cycles, which deserves more detailed quantitative scrutiny. Realistic global climate models for Jupiter must address the origins of these unexpected seasonal and non-seasonal periodicities on a virtually aseasonal Gas Giant in preparation for their eventual extension to a wider array of brown dwarfs and gas-giant planets outside our solar system. Jupiter has been used as a basis for assessing longitudinal variability in exoplanets on the basis of rotational variability[27], and variabilities have been detected on short time scales for a close-in gas giant[28] and on longer time scales for a brown dwarf[29]. The challenge of differentiating between changes arising from local meteorology, latitude-dependent dependent variations and global-scale phenomena will require time-series analyses probing timescales of days, weeks and years to unravel.

# Methods

## Basic reduction

All observations were reduced and mapped following the same procedure used by Fletcher et al.[30]. We subtracted background emission by both rapid chopping and slower nodding between Jupiter and the adjacent sky that were obtained as a part of the standard observing routine. We then applied flat-fielding corrections and geometrically calibrated the images. For the images, we projected the data onto cylindrical maps. For VISIR, due to the maximum chopping amplitude of 25 arc-seconds of the VLT and the small (32 x 32 arc-second) field of view of the instrument, the chopping-nodding technique used to remove the sky emission from the VISIR data led to a region of Jupiter being artificially obscured. This is due to part of Jupiter's thermal emission being subtracted together with the sky emission. We avoided the obscured regions from the VISIR data, resulting in only one hemisphere being available per image, which led to missing latitudes at dates where only one image was acquired.

## Filter consolidation

A wide range of instruments were used in this study, starting with scanning by single-element detectors, followed by two-dimensional imaging using MIRAC[31,32], MIRLIN[33], MIRSI[34], COMICS[35] and VISIR[36], as described in more detail in the Supplementary Information. This diversity This diversity led to images being captured by slightly different filters (as shown in Table 1S). In this study, we shifted in wavelength images captured by the different instruments to match the central wavelength of the filters in VISIR and COMICS (this is 18.72 μm and 20.50 μm) as the difference in the peak contribution functions at these wavelengths are very small. This was done by treating all images captured using filters near 18 μm and 20 μm as if they were obtained using the 18.72-μm VISIR and 20.50-μm COMICS filters, respectively, during the radiometric calibration step described below, enabling us to treat the different filters equally over the entire time series. Additionally, due to the higher diffraction-limited spatial resolutions of VISIR and COMICS observations compared to those acquired with the 3-m IRTF telescope, we smoothed VISIR and COMICS images before the calibration to match the spatial resolution of MIRAC, MIRLIN and MIRSI observations, which is used throughout the subsequent analysis.

## Radiometric calibration

Each image was radiometrically calibrated using its cylindrical-map representation by scaling the 18.72- and 20.50-μm radiances to match Voyager IRIS observations at latitudes spanning 50°S to 50°N. This was done by comparing the averaged radiance between these latitudes in each image to the averaged radiance for the same latitudes of the Voyager IRIS profile. Observations with these filters were scaled to Voyager IRIS observations because, although this wavelength range was also covered by Cassini CIRS observations in its focal plane 1, Cassini provided only hemispheric averages in this spectral region. This scaling approach was judged to be far more reliable than referencing observations of a "standard" star whose flux spectrum is known[32]. First, we made several observations on partially cloudy nights when images were acquired in transparent gaps between clouds, but observations of a stellar flux standard were not possible. Second, and more generally, our tests of this calibration approach in ostensibly clear skies yielded inconsistencies on the order of ±30%, due to variable transparency through parcels of invisible humid air in the optical path, to which these wavelengths are particularly susceptible.

**Zonally averaged radiances**

Our study examined the variability of zonal (longitudinal) mean temperatures. To achieve this, the observed radiances were averaged zonally using the mean of radiances within 30° longitude of the minimum emission angle at each latitude in 1° latitudinal bins, i.e. along the central meridian. In our long-term 5-µm radiance variability study[27], we showed that there were no substantial differences between the zonally averaged radiance using (1) this technique and (2) a second-order polynomial fitting to the emission angle.

**Smoothed radiance profiles**

We first linearly interpolated radiance profiles onto a 60-day regular grid in 1° latitude bins. We then convolved the interpolated radiance with a Savitzky-Golay (SG) smoothing filter, which fits sets of adjacent data points in a regular grid with a polynomial using a linear least-square method (https://www.l3harrisgeospatial.com/docs/SAVGOL.html). This allowed us to complete the radiance sequence when instruments were unavailable, developing radiance profiles that better represent the full dataset. This convolution was performed by linearly interpolating our radiance profiles onto a 60-day regular grid and then convolving the interpolated radiances with a 24-point wide fourth-order polynomial. Different window sizes, interpolations and polynomials were tested. Larger window sizes resulted in excessively smoothed profiles, while smaller windows and higher-order polynomials showed an artificially wavy profile. The smoothing was repeated 200 times for each latitude and wavelength, taking each time different random values of radiance within the estimated errors to consider the uncertainties of the zonally-averaged radiance. Finally, the 200 smoothed profiles were averaged together at each date to obtain an averaged smoothed radiance profile for each latitude and wavelength (see Figure 4 for an example of the radiance smoothing performed, represented by the equivalent brightness temperature). The same smoothing process was also applied to the 7.9-µm brightness temperatures shown in Fig. 1g of reference 21 that are shown in Figure 4.

**Temperature retrievals**

Temperatures were retrieved using the NEMESIS retrieval code[37] using only the 18-µm (calibrated to 18.72 µm) and 20-µm (calibrated to 20.50 µm) data interpolated onto the 60-day grid, as shown in the top and middle panels of Fig. 1, respectively. Only temperatures were allowed to vary. The chosen aerosol profile is based on an $NH_3$ ice cloud with a 10 µm ± 5 µm radius size distribution, a base at 800 mbar and top at 400 mbar, with a fractional scale height of 0.4. These values were chosen as they resulted in optimal fits to observations in a smaller data set that uses 8 filters, including those in the 7-14 µm region that are sensitive to absorption by $NH_3$ gas and aerosols. If a size distribution of aerosols with a radius of 1 µm ± 0.5 µm is assumed, the retrieved temperatures are only 0.1 K lower than for the 10-µm case on average and the temporal behavior is the same in both cases. A comparison between the temperatures retrieved from the images in all 8 filters (which includes solving for the aerosol opacity as a function of latitude and time) vs. the 18.72- and 20.50-µm images alone show differences ranging only from +1.6 K to -1.0K, well within our stated uncertainty. We note further that differences associated with variability in time, which are independent of the systematic offsets included in the differences cited above, are even smaller.

We formally derived temperatures at pressures of 100, 220 and 330 mbar, but we found that the 330-mbar temperatures were associated with the smallest uncertainties, and so we concentrated on those retrieved temperatures. The variability from a temporal mean is very similar at each pressure. Temperatures from the less frequent 1978-1982 data shown in Fig. 2 were derived from a single ~18-µm filter using a simple uniform increase or decrease of temperatures for a 'standard' vertical profile[8]. As described in Reference 8 and shown in their Fig, 2, this standard profile was derived from a smoothed average of the Voyager-1 radio-occultation results[39], with a cooler overlying stratosphere that provided a better fit to radiances observed across Jupiter's disk at all emission angles. Derived temperatures from the Voyager IRIS experiment were taken from the repository of co-author Fletcher: https://github.com/leighfletcher/Voyager, and from the Cassini CIRS experiment from a similar repository: https://github.com/leighfletcher/CassiniCIRS.

We found that using their temperatures at a pressure of 330 mbar in place of their choice of 250 mbar provided a reasonable match to the temperatures we derived for the 1983 and later, and that is what is shown in Fig, 2. A comparison of this simple approach to the derivation of 330-mbar temperatures with those resulting from the 8-filter retrievals yielded differences on the same order as our 2-filter retrievals (+1.6 K to -1.0K), also well within out stated uncertainty. One latitude region that did not appear to match well was immediately around the equator, which was much brighter (see the open circles in the upper panel of Fig. 2). We attribute this to the factor of two poorer angular resolution of the 1978-1982 scans and maps that would be 'contaminated' by the much warmer nearby North and South Equatorial Belt regions.

**Power spectra**
The power spectra shown in Figure 1 were derived from Lomb-Scargle periodograms using the IDL routine *scargle* and displayed with the IDL routine *contour*.

**Correlation analysis**
The Pearson coefficients shown in Figure 3 were derived using the IDL function *c_correlate* (https://www.l3harrisgeospatial.com/docs/c_correlate.html).

## Supplementary information

***Corresponding author. Glenn Orton. Email: glenn.orton@jpl.nasa.gov**.

**Data and materials availability:** Data are archived in: https://zenodo.org/record/7583188#.Y9dnJMHMKEY.

Acknowledgements
Some of this research was carried out at the Jet Propulsion Laboratory, California Institute of Technology, under a contract with the National Aeronautics and Space Administration (80NM0018D0004). Greco, Payne, Boydstun, and Wakefield worked on this research as interns in Caltech's Summer Undergraduate Research Fellowship (SURF) program, supported by the above funding; Payne also worked in JPL's Summer Internship Program supported by the above funds. Fletcher and Antuñano were supported by a European Research Council Consolidator Grant (under the European Union's Horizon 2020 research and innovation programme, grant agreement No 723890) at the University of Leicester. This research used the ALICE High Performance

Computing Facility at the University of Leicester. Donnelly was supported by an STFC PhD Studentship. Orton, Fletcher, Sinclair, Momary, and Yanamandra-Fisher were Visiting Astronomers at the Infrared Telescope Facility, which is operated by the University of Hawaii under contract 80HQTR19D0030 with the National Aeronautics and Space Administration. This research is based, in part, on data collected at the Subaru Telescope, which is operated by the National Astronomical Observatory of Japan; we are honored and grateful for the opportunity of observing the Universe from Maunakea, which has cultural, historial and natural significantce in Hawaii. Some of data presented herein using the Subaru Telescope were obtained by way of an exchange program with the W. M. Keck Observatory, which is operated as a scientific partnership among the California Institute of Technology, the University of California and the National Aeronautics and Space Administration. The Observatory was made possible by the generous financial support of the W. M. Keck Foundation. The authors wish to recognize and acknowledge the very significant cultural role and reverence that the summit of Maunakea has always had within the indigenous Hawaiian community. We are most fortunate to have the opportunity to conduct observations from this mountain.

We thank two anonymous reviewers for helpful and constructive comments.

**Author contributions:** GSO wrote all most of the main text and the Supplementary Information, was responsible for the general organization, led many of the observing runs and the initial reduction of the observations. AA organized the observations from the original measurements, performed the calibrations, executed the temperature retrievals and wrote a parts of the main text and the Supplemantary Information. LNF guided the spectral retrieval methodology, led many of the observing runs and the initial reduction of the observations, and helped to draft the manuscript. JAS, TWM, TF, PY-F and PTD constituted part of the teams making the observations since 2002. JJG, AVP, KAB and LEW were responsible for examining the consistency of the calibrations, testing stellar calibrations, and testing retrieval approaches on subsets of the data addressed here. All authors reviewed and commented on the manuscript. We acknowledge the work of those who made observations prior to 2002 and the authors of previous work[5,9] that inspired this study.

**Competing interests:** The authors declare no competing interests.

**Supplementary Information**

Supplementary Discussion, Figs. 1–3, Table 1 and Ref. 39.

## FIGURES

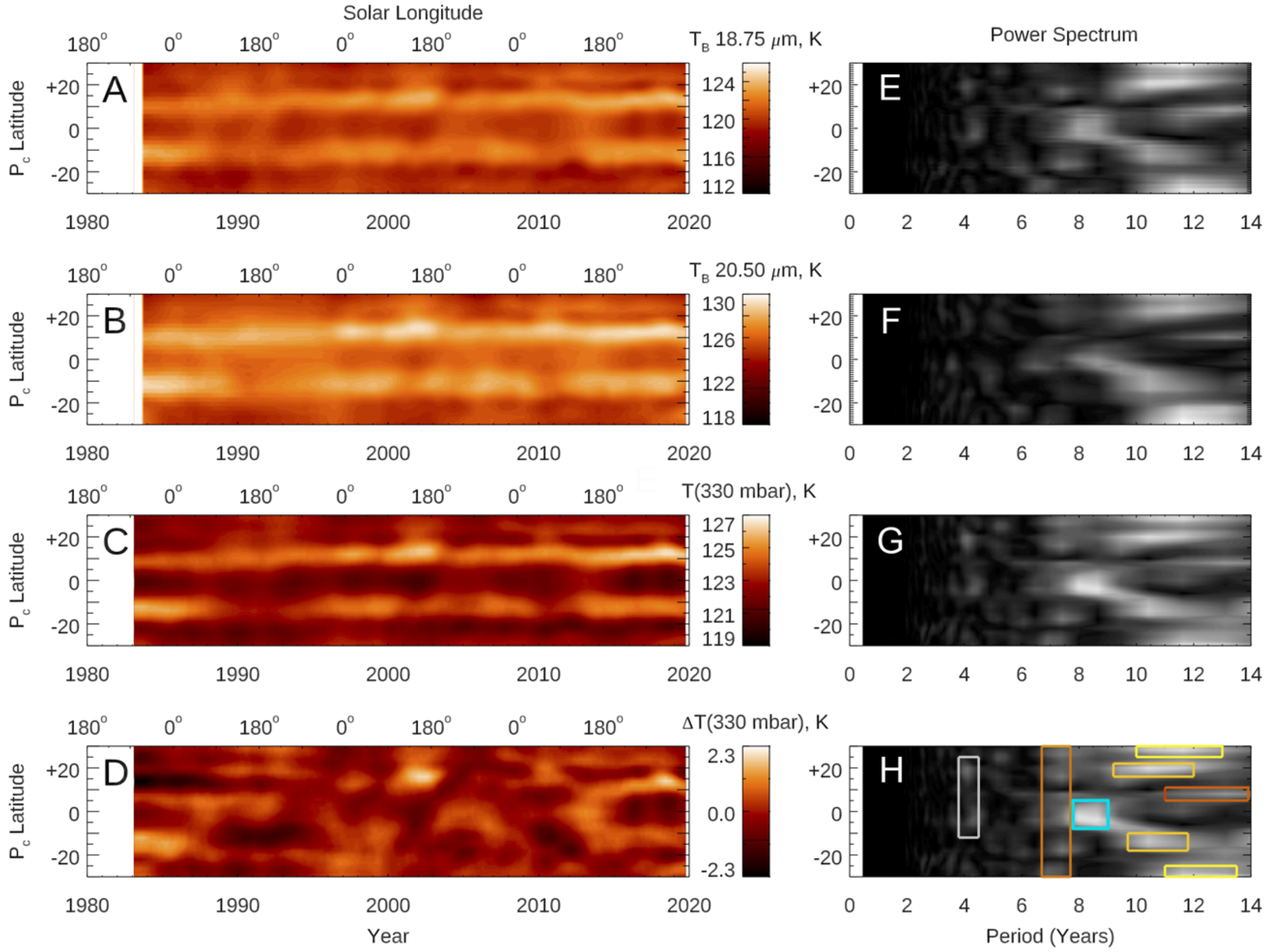

Figure 1. Brightness temperature variations along Jupiter's central meridian for two filters. a. Brightness temperatures at 18.72 µm. b. Brightness temperatures at 20.50 µm. c. 330-mbar temperatures derived from them. d. the residual 330-mbar temperatures from the zonal mean. The power spectra corresponding to these panels are shown in corresponding e-h: brightness temperature sat 18.72 µm (e), brightness temperatures at 20.50 µm (f), 330-mbar temperatures derived from them (g) and the residual 330-mbar temperatures from the zonal mean (h). The spectra were derived using Lomb-Scargle periodogram analysis, showing all results exceeding a 1% false-alarm probability. Detection of important but fainter features in the power-spectra panels is enabled by raising them to the 0.7 power, which brightens them with respect to the stronger features without enhancing noise. The power spectra in g and h are identical. h, Identifying periods that are discussed in the text. The white box bounds the latitude and period of 4.3 years with an uncertainty of ±0.5 years, the dark orange box 7.4±0.6 years, the blue box 8.3±1.0 years. The light orange boxes identify a spread of 10.4-11.7 years with an uncertainty of ±1.0 years in the north and a period of 10.5±1.0 years in the south. The yellow boxes identify a period spread of 10.3 – 12.6 ±1.5 years in the north and 11.4-13.1±1.0 years in the south. The red box identifies a broad peak period spread of 11.6-13.2 ±1.0 years. There are no substantial periods longer than 14 years.

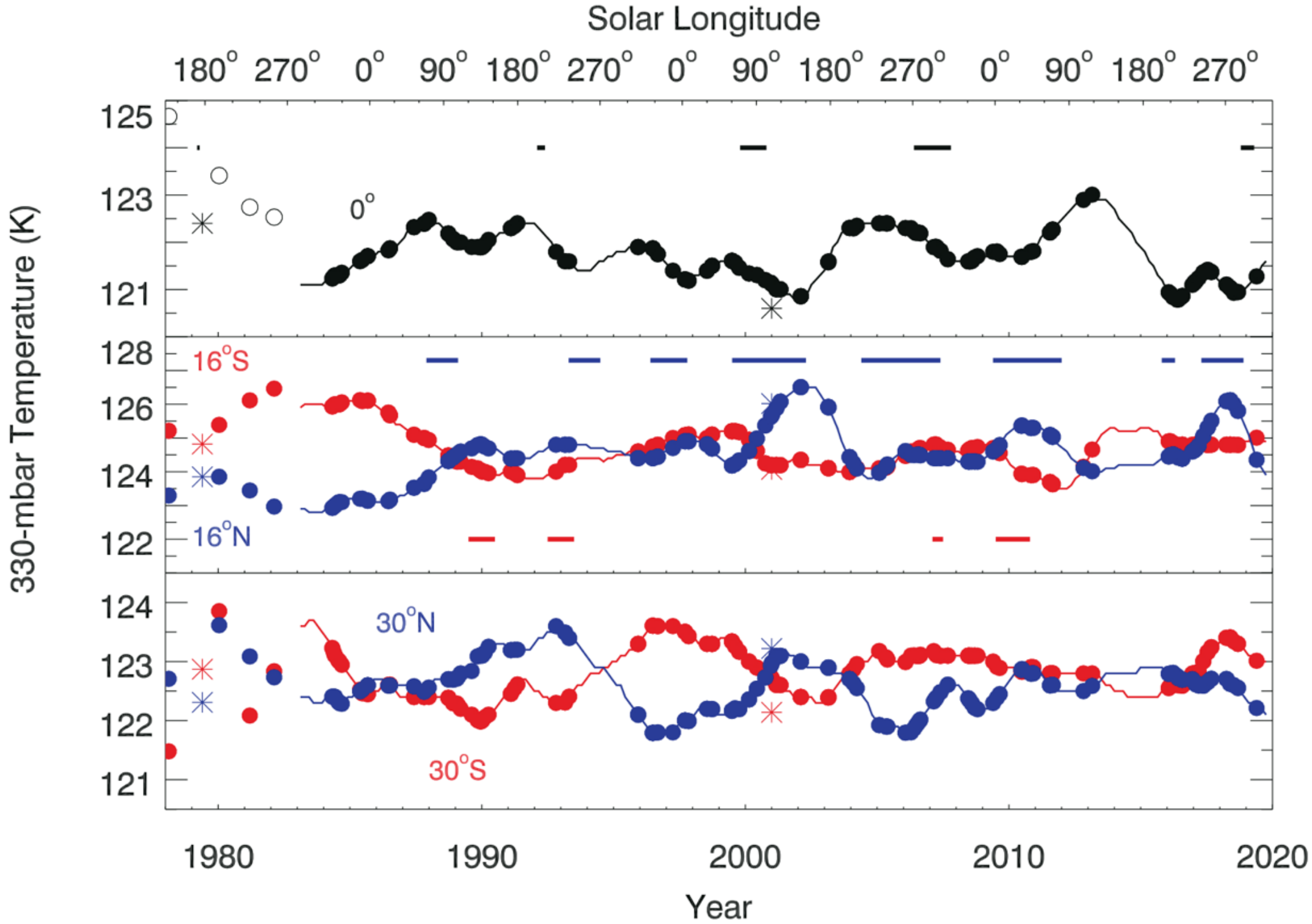

Figure 2. Retrieved temperatures at 330 mbar. a. The temperatures at the equator. b. The temperatures 16° from the equator. c. The temperatures 30° from the equator. Away from the equator, temperatures are shown at conjugate latitudes. Filled circles show the temperatures derived at each date of measurement, and solid lines indicate temperatures retrieved from 18.72- and 20.50-µm radiances interpolated on a 60-day grid. For 1978-1983, temperatures were retrieved by scaling a fixed temperature profile to match the radiance from one ~18.72-µm filter, shown by filled circles without interpolated solid lines. Open circles at the equator denote the poorer spatial resolution of these data, because their relatively high values are most likely due to contamination from the nearby brighter North and South Equatorial Belts. Asterisks denote corresponding 330-mbar temperature differences derived by the Voyager-1 IRIS instrument in 1978 (Ref. 13) and the Cassini CIRS instrument in 2001 (Ref. 14). The black horizontal bars shown with temperatures at the equator in a denote the approximate duration of Equatorial Zone disturbances[25,26]. The blue horizonal bars in b indicate the approximate duration of North Equatorial Belt expansions[10,17]. The red horizontal bars in Panel B indicate approximate duration of South Equatorial Belt fade and revival episodes[11,12]. Total formal retrieval uncertainties are 2.2K at 330 mbar, but the relative changes in time mimic those of the brightness temperatures, which are on the order of 0.2K and only slightly larger than the filled circles in this figure. For this reason and for clarity they are not illustrated by error bars.

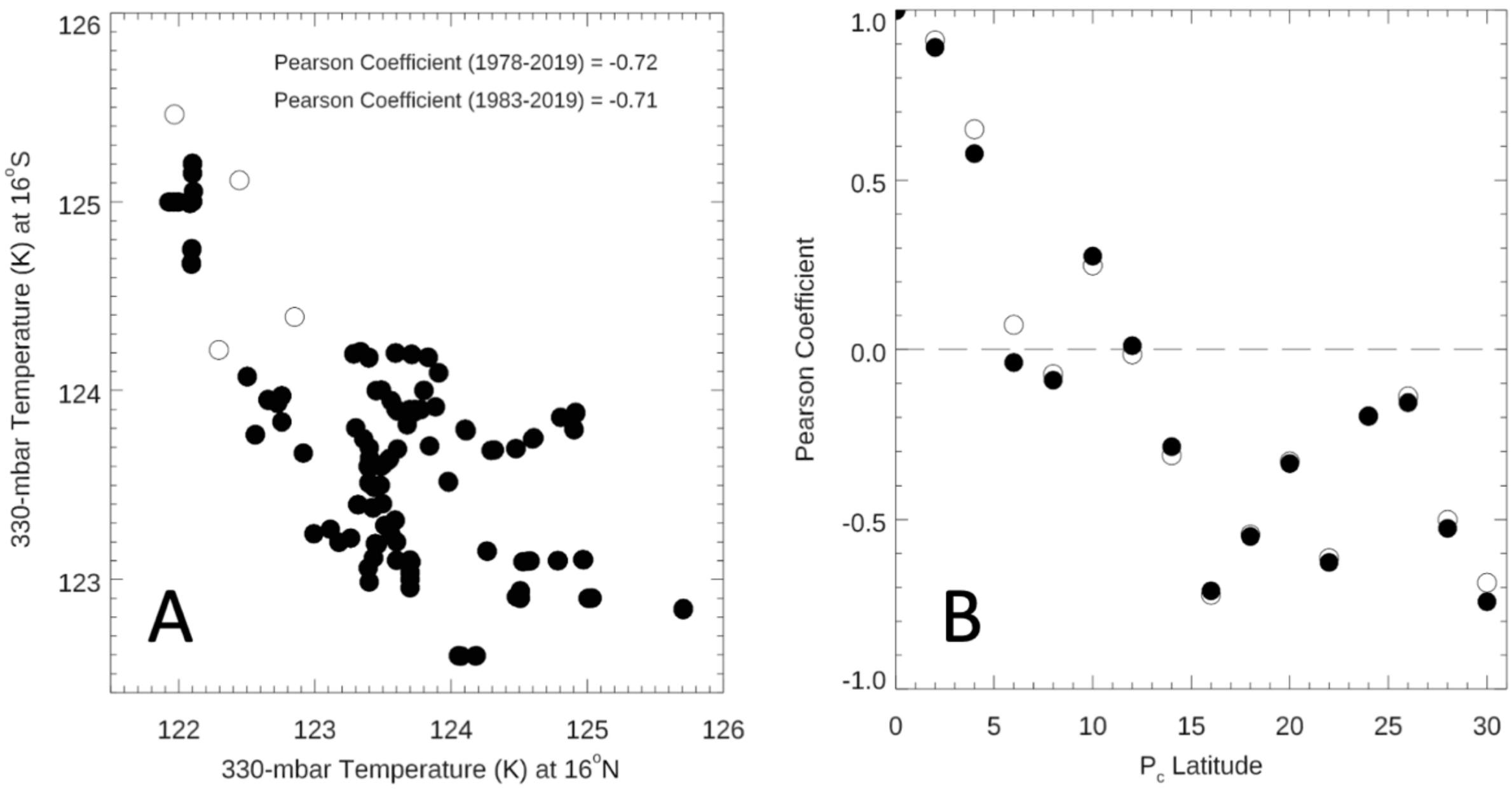

Figure 3. Inter-hemispheric correlations. a. The correlation between temperatures on all dates of observation at 16°N vs. 16°S. Filled circles represent temperatures retrieved from 1983 to 2019. Open circles represent temperatures that include those from 1978-1982, which were scaled from single-filter lower-resolution observations. The Pearson correlation coefficient is shown for both cases. As noted in Fig. 2, the formal retrieval uncertainty is 2.2K, but the point-to-point relative uncertainty is close to that of the measured brightness temperatures themselves, which is ~0.2K. b. A plot of the Pearson correlation coefficient for each latitude sampled. The highest negative values are at 16°, 22° and 30° from the equator. The strong positive values within 5° of the equator are the result of overlapping instrumental fields of view.

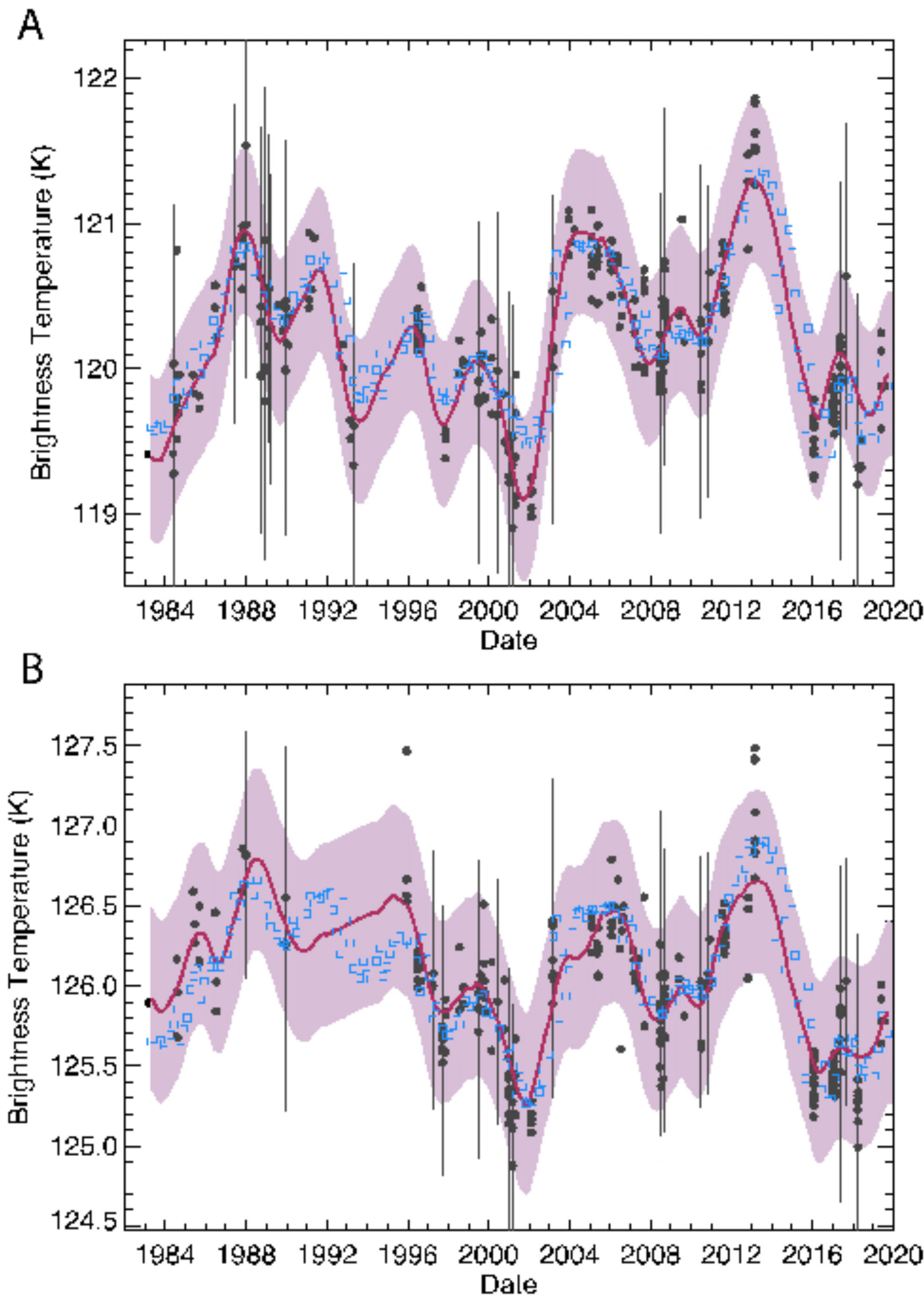

Figure 4. An example of the smoothed brightness temperature profile (solid red line) at the equator, with the purple shaded region representing its corresponding 1σ range. a,b The 18.72-μm data (a) and the 20.50-μm data (b). The smoothed brightness temperatures are compared with the raw zonal-mean brightness temperatures (black dots) used for the retrievals, together with their associated uncertainties (black error bars). The uncertainties include both calibration uncertainties and errors of the mean. The blue boxes represent the best-fit NEMESIS values to reproduce the red curve, balancing the constraints to the two filters and vertical smoothing; they all fit the red curve within the error bars.

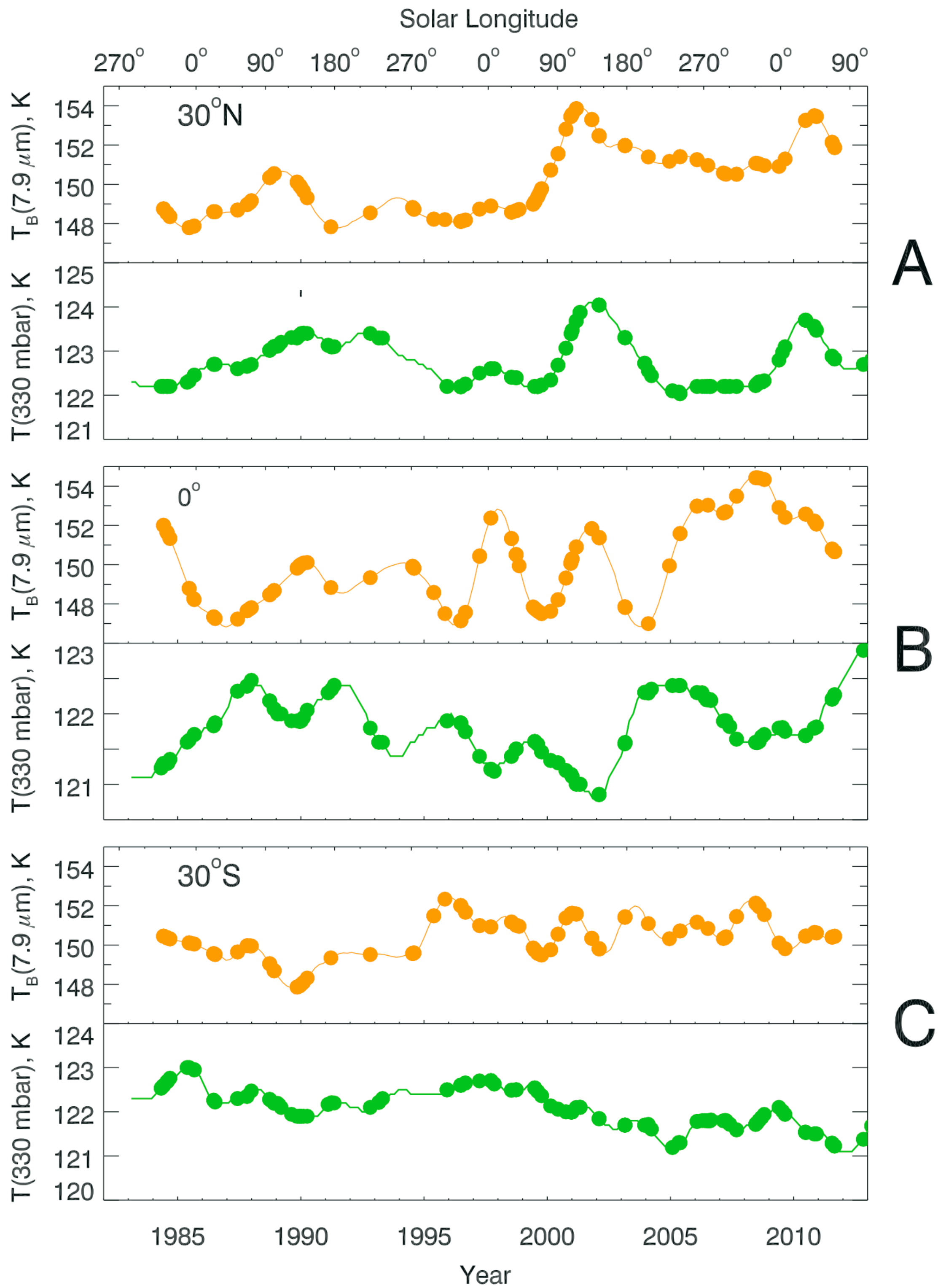

Figure 5. Comparison between variability of zonal-mean temperatures at 330 mbar (green lines and filled circles) and 7.9-µm brightness temperatures (orange lines and filled circles). a. This comparison at 30°N, b. At the equator. C, At 30°S. The 7.9-µm brightness temperatures are smoothed versions of those described by Antuñano et al.[21]